\documentclass[journal,fleqn,10pt,twocolumn,twoside]{IEEEtran}

\usepackage{amsmath,amssymb,amsfonts,bm}
\usepackage{graphicx}
\usepackage{multirow}
\usepackage{flushend}
\usepackage{cite} 
\usepackage{tikz}
\definecolor{lime}{HTML}{A6CE39}
\DeclareRobustCommand{\orcidicon}{%
	\begin{tikzpicture}
	\draw[lime, fill=lime] (0,0) 
	circle [radius=0.16] 
	node[white] {{\fontfamily{qag}\selectfont \tiny ID}};
	\draw[white, fill=white] (-0.0625,0.095) 
	circle [radius=0.007];
	\end{tikzpicture}
	\hspace{-2mm}
}
\foreach \x in {A, ..., Z}{%
	\expandafter\xdef\csname orcid\x\endcsname{\noexpand\href{https://orcid.org/\csname orcidauthor\x\endcsname}{\noexpand\orcidicon}}
}

\usepackage{hyperref} 

\begin{document}

\title{Vehicle to Vehicle Path Loss Modeling \\at Millimeter Wave Band for Crossing Cars}

\author{Anirban~Ghosh\orcidA{}, Aniruddha~Chandra\orcidB{}, Tomas~Mikulasek\orcidC{}, Ales~Prokes\orcidD{}, \\Jaroslaw~Wojtun\orcidE{}, Jan~M.~Kelner\orcidF{} and Cezary~Ziolkowski\orcidG{}
%
\thanks{This work was developed within a framework of the research grants: project no. 23-04304L sponsored by the Czech Science Foundation, MubaMilWave no. 2021/43/I/ST7/03294 funded by National Science Centre, Poland under the OPUS call in the Weave programme, and grant no. UGB/22-863/2023/WAT sponsored by the Military University of Technology.}%
\thanks{A. Ghosh is with ECE Department, SRM University AP, 522240 India.
}%
\thanks{A. Chandra is with ECE Department, NIT Durgapur, 713209 WB, India (e-mail: aniruddha.chandra@ieee.org).
}%
\thanks{T. Mikulasek and A. Prokes are with UREL, BUT, 616 00 Brno, Czechia.
}%
\thanks{J. Wojtun, J. M. Kelner and C. Ziolkowski are with ICS, MUT, 00908 Warsaw, Poland.
}%
}

\maketitle
\markboth{IEEE TRANSACTIONS ON VEHICULAR TECHNOLOGY, Vol. XX, No. X, XXX 2022}{}

\markboth{IEEE Antennas and Wireless Propagation Letters, Vol. 22, No. 9, pp. 2125-2129, September 2023, \url{https://doi.org/10.1109/LAWP.2023.3277961}}
{GHOSH \MakeLowercase{\textit{et al.}}WHEN CARS CROSS: MODELLING 60 GHz MMWAVE PATH LOSS}

\begin{abstract}
Fifth generation new radio (5G NR) is now offering sidelink capability, which allows direct vehicle-to-vehicle (V2V) communication. Millimeter wave (mmWave) enables low-latency mission-critical V2V communications, such as forward crash warning, between two vehicles crossing on a road without dividers. In this article, we present a measurement-based path loss (PL) model for V2V links operating at $59.6$ GHz mmWave when two vehicles approach from opposite sides and cross each other. Our model outperforms other existing PL models and can reliably model both approaching and departing vehicle scenarios. 
\end{abstract}


\section{Introduction}
Owing to its wideband low-latency requirements, millimeter-wave (mmWave) based fifth generation (5G) networks will be playing a significant role in vehicle-to-vehicle (V2V) communication \cite{Sakaguchi2021} in the coming years. In keeping with the advancement, 3GPP, in its release 16, introduced the standards for 5G New Radio (NR) V2X with sidelink (SL) aspect; where sidelink refers to direct communication between user equipment (UEs) without the data traveling the entire span of the network \cite{NR_V2X}. In a V2V scenario, UEs are vehicles, and direct communication between them can ensure a faster exchange of information for vehicle platooning, extended sensor data exchange, advanced driving, remote driving, and intent sharing. For example, the NR cellular-V2X sidelink would improve reliability in autonomous driving through services such as forward crash warning \cite{qualcomm}. 

When two cars are approaching each other from the opposite direction, there are multiple applications that require data exchange between them. This includes sharing information about the road condition, traffic situation, or diversion, as well as emergency warnings like collision avoidance between approaching vehicles. Communication can start between vehicles moving in opposite directions only when they are within communication range. As the cars are speeding from opposite sides, the relative velocity is additive, and there is a very small window for communication. Thus, we require a high bandwidth link to effectively transfer data between them and take actions as deemed fit. 


V2V channel sounding campaigns in various scenarios (urban, suburban, parking garage, highway, intersections) were primarily restricted to frequencies below 6 GHz \cite{boban,ding,haneda2, bjtu1,mexico1}. In \cite{garcia} the authors performed a measurement campaign using directional antennas both at the transmitter and the receiver, and fixed them at a low elevation position on the car bumpers for V2V communications at 38 GHz and 60 GHz frequency bands. In most practical applications, it makes more sense to place the antennas on the roof of the vehicles to reduce attenuation and path loss (PL). In addition, the measurement was carried out in a convoy configuration. In \cite{dupleich}, the measurement was carried out at $60$ GHz, but here the transmitter and receiver were placed on a fixed tripod emulating a rooftop antenna of a stranded vehicle. Besides, here also the configuration explored was vehicles moving away from one another and not approaching. As an interesting study, it was pointed out in \cite{park} that PL can be different for varying antenna positions, even at $28$ GHz. However, there was no study on the PL model to see if the PL, when cars are approaching, should be different from the case when the vehicles are moving away, and if this changes with relative speeds.
\begin{figure*}[ht!]
      \centering 
      \includegraphics[width=0.8\textwidth]{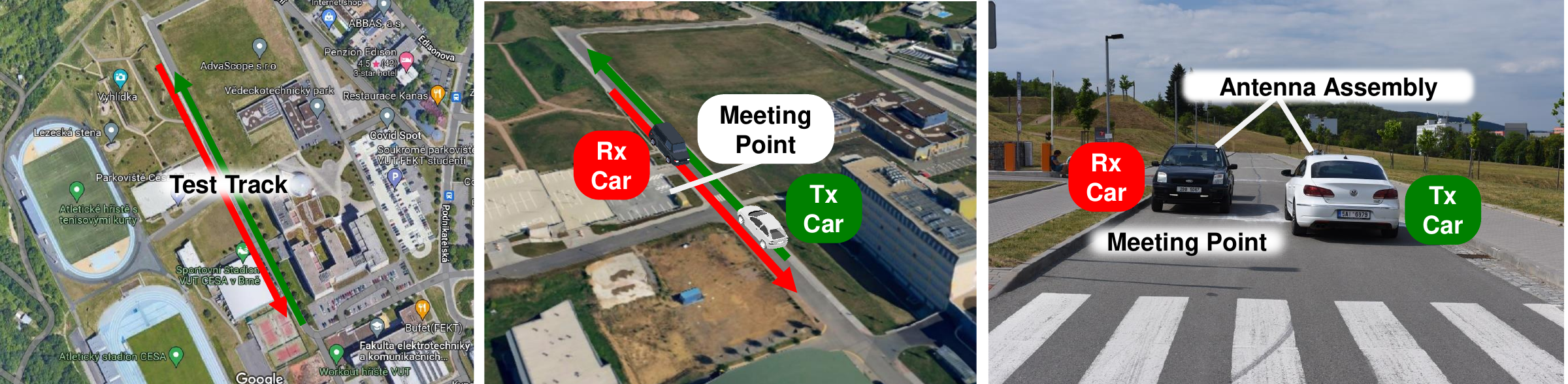}
      \caption{(left) The aerial view of the test site [courtesy: Google Maps], (middle) the 3D site layout and measurement tracks [courtesy: Mapy.cz], and (right) photograph showing the vehicles under test and antenna hardware assembly.}
      \label{fig:measurement}
\end{figure*}

Thus to address the aforementioned gap, we performed a measurement campaign at $59.6$ GHz in a V2V
communication scenario \cite{prokes,prokes2}. 
The current paper focuses on the PL values obtained from the field tests; specifically:
\begin{itemize}
    \item Based on real-world data, a PL model is proposed for a typical mmWave band ($59.6$ GHz) V2V link where vehicles approach each other, cross, and move away. 
    
    \item To quantize the goodness of fit of our introduced model we compare it with four other existing PL models in terms of root mean square error (RMSE), Grey relational grade-mean absolute percentage error (GRG-MAPE), and Pearson correlation coefficient-mean absolute percentage error (PCC-MAPE). Our model showed a better fit to experimental data with respect to other standard PL models.
    
    \item The proposed model is parameterized to best fit the measurement data. The parameterized model confirms our intuition that the same model with a different parameter set is required for defining each scenario (moving in and moving away) at a different relative speed.
\end{itemize}

\section{Measurement Campaign}
The transmitter (Tx) of the time-domain channel sounder as described in \cite{prokes}, comprises of Anritsu MP1800A signal quality analyzer that supplies the seamlessly repeating Golay complementary sequence at baseband (BB) with a bandwidth of $4$ GHz. The BB signal is passed through an optional low pass filter (LPF) to maintain bandwidth integrity before being upconverted into the center frequency of $59.6$ GHz using a SiversIma FC1005V/00 V-band up/down converter. To compensate for phase noise, the reference signal for up-conversion is applied from an Agilent E8257B frequency-stable, low-phase noise generator. The output signal from the converter passes through a band pass filter (BPF) for image frequency rejection and a power amplifier (PA) of gain $35$ dB to indemnify for propagation loss. The amplified signal is then transmitted using an omnidirectional substrate-integrated waveguide (SIW) antenna. 

At the receiver (Rx), a similar antenna is used for the reception of the transmitted signal. The received signal initially passed through a low noise amplifier of gain $33$ dB for loss compensation before being mixed with the signal from a carrier generator Agilent 83752A inside SiversIma FC1003V/01 V-band up/down converter to obtain the baseband signal \cite{prokes}. The obtained baseband signal is stored using Tektronix 72004C ($20$ GHz, $50$ GS/s) mixed signal oscilloscope (MSO). Downloading data from the oscilloscope and subsequent basic processing are facilitated by LabView software. The synchronization between the Tx and Rx module is established through back-to-back (B2B) calibration. Post calibration, the synchronization is maintained with the help of rubidium (Rb) oscillators. The transmitted Golay sequence is generated at a data rate of $R_\mathrm{DR}=12.5$ Gbps and contains $N=2048$ bits, and hence the maximum observable time is $T_\mathrm{max}=N/R_\mathrm{DR}=163.8$ ns. With a memory depth of the MSO at $M_\mathrm{D}=31.25$ MSample/channel, the sampling rate at $R_\mathrm{S}=50$ GSample/s, and the sampling interval at $T_\mathrm{int}=5$ ms, the number of samples per channel impulse response (CIR), $N_\mathrm{sam/CIR}=N \times R_\mathrm{S} / R_\mathrm{DR}$, the number CIRs per measurement, $N_\mathrm{CIR/meas}= M_\mathrm{D} / (8\times N_\mathrm{sam/CIR})$ and total time for each measurement, $T_\mathrm{meas}=N_\mathrm{CIR/meas}\times T_\mathrm{int}$ are respectively $8192$, $468$ and $2.34$ ms.


The V2V measurement campaign \cite{prokes2} was conducted on the campus of the Brno University of Technology in Brno, Czech Republic. In Fig.~\ref{fig:measurement} we provide a bird's eye view of the test site. No other cars except for the participating ones, nor any notable moving objects were present during the measurements. There are no buildings around the immediate perimeter of the driven two-lane road except for the ones as shown in \ref{fig:measurement}. The location and controlled traffic ensure the availability of a non-obstructed line-of-sight (LOS) throughout the measurement campaign. 
A white Volkswagen Passat CC 2.0 TDI car houses the antenna assembly for the Tx, and a black Ford Fusion 1.4i contains the Rx antenna and associated circuitry. While the uninterrupted power supply (UPS) was enough to power the sequence generator in the Tx section, the MSO drained the UPS within minutes. This is why a trailer (not seen in the photograph) is attached at the back of the Rx car with an additional power supply.

\begin{figure}[ht!]
      \centering 
      \includegraphics[width=0.8\columnwidth]{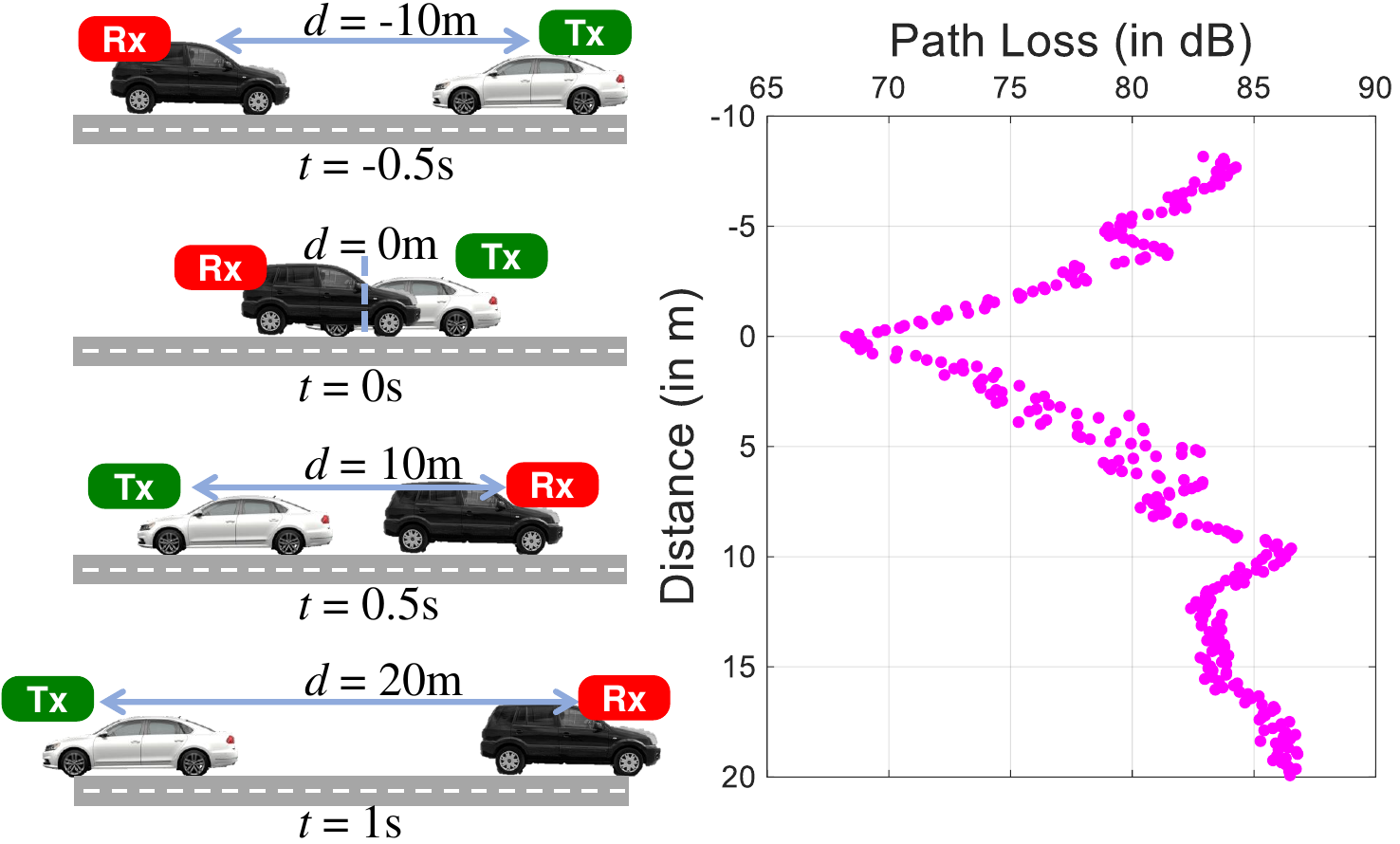}
      \caption{Typical variation of PL values with the relative distance between the cars, when cars cross along the test track. Zero distance means the cars are at the meeting point, a positive value of relative distance denotes cars are departing and a negative value denotes they are approaching each other. Relative speed is $70$ km/h ($\sim20$ m/s).}
      \label{fig:pathloss}
\end{figure}
The Tx and Rx antennas of the sounder were placed using suction caps on the driver's side of the respective vehicles. The measurement is carried out for six passes of the vehicles at two different relative speeds - $4$ passes at $50$ km/h ($13.89$ m/s), and $2$ passes at $70$ km/h ($19.44$ m/s). The separation between Tx and Rx at the beginning of each pass is approximately $35$ m. As is expected, due to varying relative speeds the rendezvous point of the Tx and Rx keeps varying with respect to the starting point of the Tx. A typical crossing on the measurement track generates a PL profile as shown in Fig.~\ref{fig:pathloss}.
\section{PL Models}
Next, we introduce four standard PL models developed for different urban scenarios such as Urban Microcell (UMi) and Urban Macrocell (UMa) \cite{3gpp1} which will be used as the basis for comparison.

{\bf Floating intercept (FI):}
FI PL model \cite{fimodel} is defined as
\begin{equation}
	    PL(d) = \alpha + 10\beta\log_{10}(d) + \mathcal{X}_{\sigma} \label{eqn:fi}
\end{equation}
where $\alpha$ is the floating intercept, $\beta$ is the PL exponent (PLE), and $\mathcal{X}_{\sigma}$ is a zero mean Gaussian random variable representing the large scale shadow fading with standard deviation $\sigma$ (in dB). In \cite{fimodel}, for UMi with different directional antennas the PLE and $\sigma$ were in the range $0.4-4.5$ and $5.78-8.52$ dB respectively depending on the Tx height when the Tx-Rx separation was varied over $30-200$ m. The frequencies of interest in the work were $28$ and $38$ GHz.

{\bf Close-in (CI):}
The CI PL model \cite{cimodel} is given as 
\begin{equation}
		PL(f,d) = FSPL(f, d_0) + 10\beta\log_{10}\Big(\frac{d}{d_0}\Big) + \mathcal{X}_{\sigma} \label{eqn:ci}
\end{equation}
where $d_0 = 1$ m, $f$ is the center frequency in Hz, and free space PL (FSPL) is a frequency-dependent parameter, $FSPL(f, d_0) = 20\log_{10}\left(4\pi f/c\right)$, where $c$ is the speed of light. FSPL can be calculated as $67.95$ dB for our measurement center frequency of $59.6$ GHz. In \cite{cimodel} it was observed that for frequencies upto $100$ GHz the $\beta$ and $\sigma$ values varied between $1.85-1.98$ and $3.1-4.2$ dB respectively for UMi and were $2$ and $4.1$ dB respectively for UMa.

{\bf alpha-beta-gamma (ABG):}
The third model is the alpha-beta-gamma PL model \cite{abgmodel}
\begin{equation}
PL(d) = \alpha + 10\beta\log_{10}\Big(\frac{d}{d_0}\Big)  
+ 10\gamma\log_{10}\Big(\frac{f_c}{f_0}\Big) + \mathcal{X}_{\sigma}\label{eqn:abg}
\end{equation}
where $f_0 = 1$ GHz and $\gamma$ denotes a frequency dependent PLE. The PL model parameters, $\alpha=2.1$, $\beta=31.7$ dB, $\gamma=2$ and $\sigma=3.9$ dB were obtained in \cite{abgmodel} from the extensive campaign at different frequencies below $60$ GHz with varying Tx-Rx separation in UMa.

{\bf 3GPP:} 
The fourth model is the PL model proposed for V2V communication by 3GPP in its release 15 for LOS communication in an urban setting \cite{3gpp.36.331}. The model is described as follows
\begin{equation}
           PL_{\mathrm{LOS}}^{\mathrm{urban}}(d) = 38.77 + 16.7\log_{10}(d) + 18.2\log_{10}(f_c) + \mathcal{X}_a \label{eqn:3gpp}
\end{equation}
where the shadowing, i.e. the effect of signal power fluctuations due to surrounding objects is modeled as lognormal distribution with standard deviation $\mathcal{X}_{\sigma} = 3$ dB. 

In this context, it is to be noted that the parameters delineated from the cited references of each model in this section are only for LOS scenarios which is the scenario of interest in the current campaign.

\section{Proposed PL Model and Comparison with other PL Models}
The measurement sets for different relative velocities are segregated into two scenarios - one, when the Tx is moving towards the Rx, termed as \textit{moving in scenario} and the one when they are \textit{moving away} from one another from the meeting point. 

We propose a PL model for both cases in line with the 3GPP model with a few adaptations as follows
\begin{equation}
		PL(d) = \eta_1 + 18.2\log_{10}(f_c) + \eta_2\log_{10}(d) + \mathcal{X}_{\sigma} \label{eqn:th_model}
\end{equation}
The modeled parameters - constant term, $\eta_1$, and log distance dependent coefficient, $\eta_2$ are obtained by fitting our model with the measured data set for various runs. For the shadowing term, $\mathcal{X}_{\sigma}$, the standard deviation $(\sigma)$ varies with relative speed and direction of motion and can be derived as, $\sigma = \sqrt{ \left(\sum\nolimits_{i=1}^{N}|x_i - \Bar{x}|^2 \right)/(N - 1)} $ \cite{rappaport}, where $x_i$ are the sample values, $\Bar{x}$ is the mean of the samples and $N$ is the total number of samples. 

\begin{figure}[ht!]
      \centering 
      \includegraphics[width=\columnwidth]{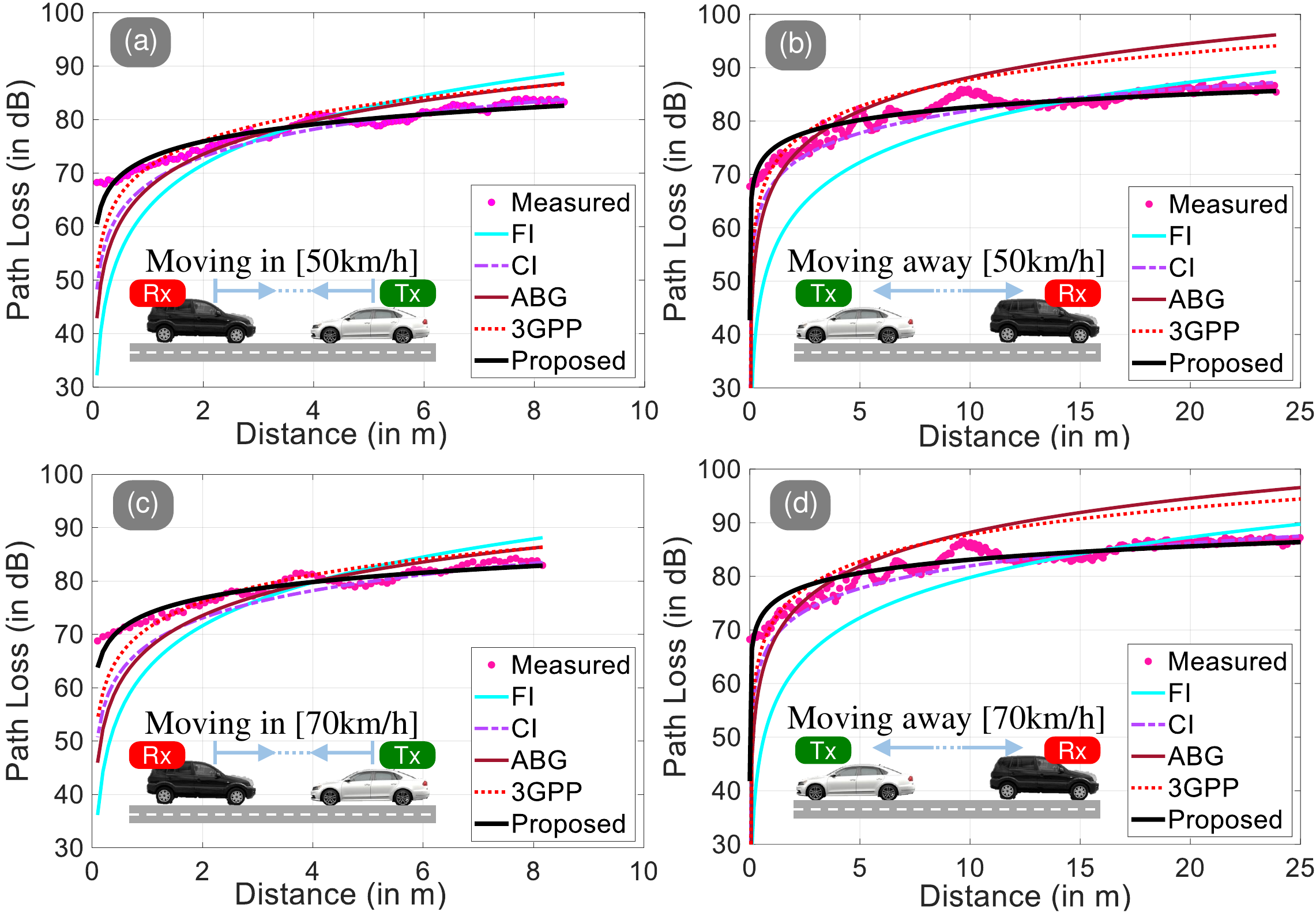}
      \caption{Comparison of PL models for different relative velocities: [a,b] $50$ km/h, [c,d] $70$ km/h.}
      \label{fig:fig3}
\end{figure}
For a visual comparison, the measured data, along with the proposed model and the four standard PL models described in the previous section, are plotted in Fig.~\ref{fig:fig3} when the relative speed is $50$ km/h and $70$ km/h. The parameters used for the standard models are chosen so as to obtain the best goodness of fit (GoF) with the measurement data. While the cumulative distance span for \textit{moving in} and \textit{moving away} scenarios in Fig. \ref{fig:fig3} remains the same ($\sim 35$ m), the rendezvous point varies with relative speed. This can be verified from the difference in the span of Tx-Rx separation, i.e., the horizontal axis in Fig. \ref{fig:fig3}. On the other hand, a similar trend in measurement data for \textit{moving in} (Fig. \ref{fig:fig3} (a) and \ref{fig:fig3} (c)) or \textit{moving away} (Fig. \ref{fig:fig3} (b) and \ref{fig:fig3} (d)) confirms the stationarity of the measurement environment. The figures clearly demonstrate that the proposed model is a better fit compared to the standard models.

\begin{table}[ht]
\caption{GoF measure of PL models for different relative speeds and direction of motion}
\centering
\begin{tabular}{|p{1.2 cm}|p{1.3 cm}|l|c|c|c|}  \hline
{\bf Direction} & {\bf Relative} & {\bf PL}    & {\bf RMSE} & {\bf GRG}  & {\bf PCC}  \\
{\bf of motion} & {\bf Speed}    & {\bf Model} &            & {\bf MAPE} & {\bf MAPE} \\ \hline
\multirow{10}{1.2 cm}{Moving in} & \multirow{5}{1.5 cm}{70 km/h 19.44 m/s} 
   & FI       & 5.36 & 0.93 & 0.95 \\ \cline{3-6}
 & & CI       & 3.92 & 0.96 & 0.96 \\ \cline{3-6}
 & & ABG      & 4.70 & 0.94 & 0.95 \\ \cline{3-6}
 & & 3GPP     & 3.02 & 0.96 & 0.97 \\ \cline{3-6}
 & & Proposed & {\bf 1.22} & {\bf 0.98} & {\bf 0.98} \\ \cline{2-6}
 & \multirow{5}{1.5 cm}{50 km/h 13.89 m/s}  
   & FI       & 5.23 & 0.94 & 0.95 \\ \cline{3-6}
 & & CI       & 3.45 & 0.96 & 0.97 \\ \cline{3-6}
 & & ABG      & 4.44 & 0.95 & 0.96 \\ \cline{3-6}
 & & 3GPP     & 3.18 & 0.96 & 0.96 \\ \cline{3-6}
 & & Proposed & {\bf 1.36} & {\bf 0.98} & {\bf 0.98} \\ \hline
 \multirow{10}{1.2 cm}{Moving away} & \multirow{5}{1.5 cm}{70 km/h 19.44 m/s} 
   & FI       & 7.64 & 0.94 & 0.94 \\ \cline{3-6}
 & & CI       & 3.48 & 0.98 & 0.97 \\ \cline{3-6}
 & & ABG      & 9.43 & 0.91 & 0.91 \\ \cline{3-6}
 & & 3GPP     & 7.38 & 0.93 & 0.92 \\ \cline{3-6}
 & & Proposed & {\bf 1.96} & {\bf 0.99} & {\bf 0.98} \\ \cline{2-6}
 & \multirow{5}{1.5 cm}{50 km/h 13.89 m/s}  
   & FI       & 8.84 & 0.93 & 0.93 \\ \cline{3-6}
 & & CI       & 3.77 & 0.97 & 0.96 \\ \cline{3-6}
 & & ABG      & 8.28 & 0.92 & 0.92 \\ \cline{3-6}
 & & 3GPP     & 6.64 & 0.94 & 0.93 \\ \cline{3-6}
 & & Proposed & {\bf 2.05} & {\bf 0.98} & {\bf 0.97} \\ \hline
\end{tabular}
\label{Table:1}
\end{table}
Table~\ref{Table:1} shows how closely the various PL models fit our measurement dataset in terms of root mean square error (RMSE), Grey Relational Grade-mean absolute percentage error (GRG-MAPE), and Pearson correlation coefficient-mean absolute percentage error (PCC-MAPE) \cite{Yu}. According to Grey system theory the GRG ranges from $0$ to $1$ which estimates the matching degree between the measured data and that estimated from a typical model. MAPE on the other hand differs from GRG and can provide another dimension in model selection to match the measured data. When both the methods are combined the accuracy of the model selection method can be improved and is given as 
\begin{equation}
		\rho_{\mathrm{grg - mape}} = |\alpha.\rho_{\mathrm{grg}} + \beta. \rho_{\mathrm{mape}}| 
\end{equation}
where $\rho_{\mathrm{grg}}$ gives the similarity measure using solely the GRG algorithm proposed in \cite{Yu}. The parameters, $\alpha$ and $\beta$, are the respective weighting factors for GRG and MAPE whereas $\rho_{\mathrm{mape}}$ is given as
\begin{equation}
		\rho_{\mathrm{mape}} = |1 - Per_{e}| = |1 - \frac{1}{n}\sum_{k = 1}^{n}\Big(\frac{|x_{i}(k) - x_{0}(k)|}{x_{0}(k)}\Big)| 
\end{equation}
where in our case $x_0$ denotes measurement data, $x_i$ denotes predicted value using a PL model and $n$ represents the total number of data points. PCC on the other hand as proposed by K. Pearson can evaluate the correlation between two data sets and produces a value in the range of $-1$ to $1$ depending on negative or positive correlation respectively. The algorithm, 
however, can be modified to produce values in the range $0$ to $1$ for the ease of comparison with other similarity measures such as GRG. For the same reason as GRG - MAPE, PCC can be combined with MAPE to improve the accuracy of the algorithm producing
\begin{equation}
		\rho_{\mathrm{pcc - mape}} = |\alpha.\rho_{\mathrm{pcc}} + \beta. \rho_{\mathrm{mape}}| 
\end{equation}

In each case, the PL value obtained from the model under consideration is compared with the measurement dataset to calculate the RMSE value and also GRG - MAPE and PCC - MAPE values with a weighting factor, $\alpha$ of $0.1$ for GRG or PCC and a weighting factor, $\beta$ of $0.9$ for MAPE (or error's effect). The different PL model selection methods for channel modeling as seen from Table~\ref{Table:1} unanimously point to the fact that the FI model fits the worst for the dataset being considered while our proposed model quite closely captures the PL pattern of the dataset. The poor performance of the FI model, in this case, can be attributed to the lack of presence of any frequency-dependent term in the model unlike in the proposed or other considered models.

\section{PL Models for Moving In/ Moving Away}
\begin{figure}[ht!]
      \centering 
      \includegraphics[width=\columnwidth]{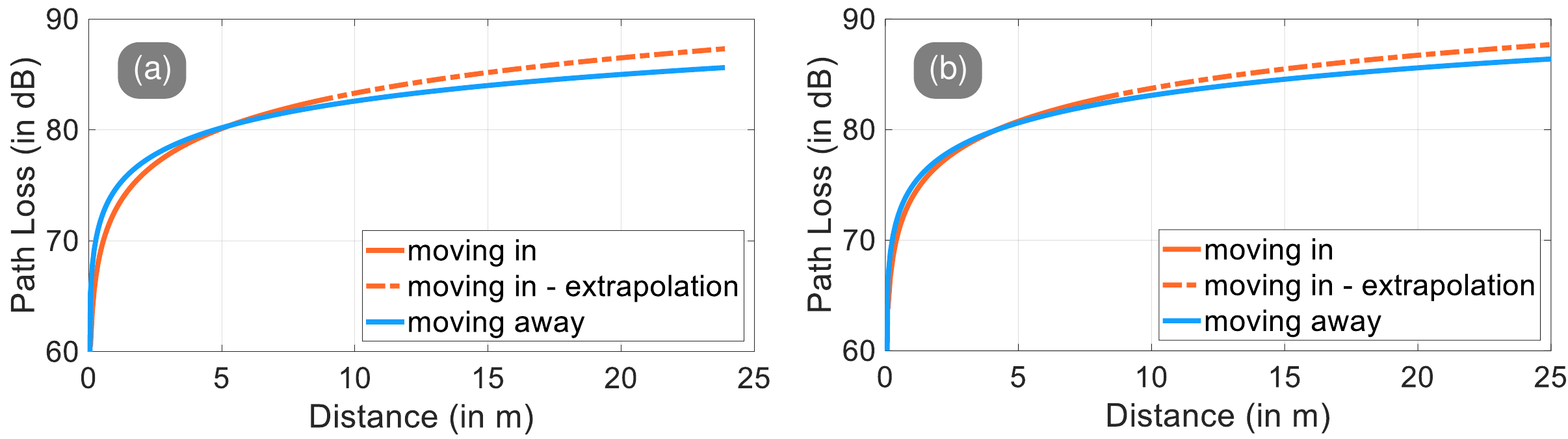}
      \caption{Comparison of proposed PL model for \textit{moving in} and \textit{moving away} scenarios: (a) Relative speed $=50$ km/h, (b) Relative speed $=70$ km/h.}
      \label{fig:movinaway}
\end{figure}

The proposed models are also compared for the two scenarios of interest to us, i.e \textit{moving in} and \textit{moving away} as can be seen in Fig. \ref{fig:movinaway}. The extrapolation refers to the region predicted by the proposed model only (no measured data available for verification).

It is seen that irrespective of the relative speed, the PL is more for departing vehicles when the separation between Tx and Rx is less than $\sim 5$m and the trend reverses as the separation increases. The crossover in the trajectory happens at a separation of approximately $4.12$ m when the relative speed is $70$ km/h and at a separation of around $5.2$ m when the relative speed is $50$ km/h. The maximum difference in PL in the two scenarios, however, never exceeds $2$ dB irrespective of the relative speed. 

Parameters for the proposed model for different directions of motion and varying relative speed are summarised in Table \ref{Table:2}.
\begin{table}[ht]
\caption{Proposed model parameters for different directions of motion and varying relative speed}
\centering
\begin{tabular}{|p{1.2 cm}|p{1.3 cm}|c|c|c|}  \hline
{\bf Direction} & {\bf Relative} & $\bm{\eta_1}$ & $\bm{\eta_2}$ & $\bm{\sigma}$ \\
{\bf of motion} & {\bf Speed}    &               &               & {\bf (in dB)} \\ \hline
\multirow{2}{1.2 cm}{Moving \\in} & $50$ km/h & 40.42 & 10.59 & 6.34 \\ \cline{2-5}
                                  & $70$ km/h & 41.51 &  9.92 & 5.98 \\ \hline
\multirow{2}{1.2 cm}{Moving away} & $50$ km/h & 42.31 &  7.99 & 5.99 \\ \cline{2-5}
                                  & $70$ km/h & 42.53 &  8.26 & 6.13 \\ \hline
\end{tabular}
\label{Table:2}
\end{table}
The values indicate that the effect of the direction of motion is more prominent than the relative speed. In fact, with a little loss of accuracy, it is possible to derive an average model for two scenarios irrespective of speed. The \textit{moving in} conditions can be characterized with $\eta_1\approx41,\eta_2\approx10$, and for \textit{moving away}, the parameters would be $\eta_1\approx42,\eta_2\approx8$. The standard deviation of shadow fading can be considered fixed at $6$ dB for all cases.

\section{Conclusions}
The current study presents a measurement-validated PL model for V2V scenarios - approaching and departing vehicles. The scenarios under investigation assume great importance in implementing automated collision avoidance and facilitating V2X with SL. The performance of the proposed model is assessed with other standard PL models and is found to outperform them in terms of various metrics. Furthermore, in the context of vehicles approaching, crossing, and eventually departing, the model is found to be dependent more on the scenario than the relative speed. 


\bibliographystyle{./IEEEtran}    
\bibliography{./Ref}
\end{document}